\documentclass[aps,pre,reprint,groupedaddress,twocolumn]{revtex4-2}
\usepackage[utf8]{inputenc}
\usepackage{graphics,graphicx,amssymb,amsmath,dcolumn,bm,url,amsbsy,pgf}
\usepackage{booktabs}
\usepackage{geometry}
\usepackage{indentfirst}
\usepackage{CJK}
\usepackage{url}

\begin{document}

\title{A nontwist field line mapping in a tokamak with ergodic magnetic limiter}
\author{Michele Mugnaine}
\email[]{mmugnaine@gmail.com}
\affiliation{Institute of Physics, University of S\~ao Paulo, S\~ao Paulo, SP, Brazil}
\author{Jos\'e D. Szezech Jr}
\affiliation{Postgraduate Program in Sciences, State University of Ponta Grossa,  Ponta Grossa, PR, Brazil}
\affiliation{Department of Mathematics and Statistics, State University of Ponta Grossa, Ponta Grossa, PR, Brazil}
\author{Ricardo L. Viana}
\affiliation{Department of Physics, Federal University of Paran\'a, Curitiba, PR, Brazil}
\author{Iber\^e L. Caldas}%
\affiliation{Institute of Physics, University of S\~ao Paulo, S\~ao Paulo, SP, Brazil}

\date{\today}
\begin{abstract}
For tokamaks with uniform magnetic shear, Martin and Taylor have proposed a symplectic map has been used to describe the magnetic field lines at the plasma edge perturbed by an ergodic magnetic limiter. We propose an analytical magnetic field line map, based on the Martin-Taylor map, for a tokamak with arbitrary safety factor profile. With the inclusion of a non-monotonic profile, we obtain a nontwist map which presents the characteristic properties of degenerate systems, as the twin islands scenario, the shearless curve and separatrix reconnection. We estimate the width of the islands and describe their changes of shape for large values of the limiter current. From our numerical simulations about the shearless curve, we show that its position and aspect depend on the control parameters.
\end{abstract}
\maketitle

\section{Introduction}

Tokamaks are toroidal devices built with the purpose to confine plasma magnetically.  This confinement is required since the plasmas generated by ohmic heating of the filling gas present such high temperatures that no material could withstand them \cite{wesson}. These magnetically confined toroidal plasmas are promising candidates for obtaining a feasible process of fusion energy generation, since they present potential for succeeding in stable and operating conditions composed of high plasma pressures and long confinement times \cite{evans}. The confinement in tokamaks is a limiting factor for the effectiveness of obtaining energy by thermonuclear fusion. For this reason, modifications and new procedures involving tokamak design are  constantly being developed in order to improve the confinement. A widely used method to provide such improvement is the creation of chaotic magnetic field lines in the periphery of the plasma, which results in  the destruction of the flux surfaces, the plasma edge cooling, and more control over the plasma-wall interactions \cite{evans,portela2003,martin1984}. 

The Ergodic Magnetic Limiter (EML), idealized by Karger and Lackner \cite{karger1977}, is composed by a set of current-carrying wires externally placed in the Tokamak \cite{martin1984,elton2002}. The EML generates a magnetic field which resonates with the equilibrium fields of the plasma, leading to the emergence of magnetic islands, which eventually overlap and result in the magnetic surfaces destruction and in the formation of a chaotic layer at the edge of the plasma \cite{martin1984,elton2002,mccool1989}. This chaotic layer, also called ergodic or stochastic, is composed by area-filling field lines which are expected to decrease heat and particle loading on the tokamak wall, restricting the level of plasma contamination by impurities released from the wall \cite{martin1984}. {\color{black} The inclusion of ergodic limiter has already been performed for some machines, as the TCABR (Tokamak Chauffage Alfvén Brésilien)\cite{elton2002,pires2005}, the original TEXT (Texas Experimental Tokamak) \cite{mccool1989,ohyabu1984}, HYBTOK-II \cite{shen1989} and JFT-23 \cite{tamai1995}, to name a few. }

As in other examples in plasma physics, presented and discussed in Refs. \cite{evans,escande2016,escande2018}, the effect of the ergodic limiter on the magnetic field lines in tokamaks can be treated by the Hamiltonian formalism and the resulting dynamics can be investigated by two dimensional symplectic maps. In order to illustrate general aspects of the EML, Martin and Taylor \cite{martin1984}  proposed a two-dimensional area-preserving map that describes the role of the rotational transform, the shear, the diffuser strength, periodicity and length for a scenario of tokamaks with EML. In the analytical derivation, the effect of the limiter is considered to be restricted to a small part of the toroidal circumference, where the magnetic shear can be neglected, and the map is appropriate to describe the region near the tokamak wall  \cite{martin1984}. The Martin-Taylor model was revisited in Ref. \cite{portela2003}, in which the authors discuss some results involving periodic orbits, the position of the resonance islands, the emergence of global chaos, diffusion and the size of the chaotic region. In \cite{portela2003}, the authors were able to find approximate results for the map when considering the scenario in which the perturbation is small and the map can be treated as a near-integrable system.

The Martin-Taylor map describes the field line dynamics in a region close to the tokamak wall, for scenarios with an uniform magnetic shear and a safety factor radial profile increasing monotonically from the magnetic axis to the tokamak wall \cite{portela2003}. However, the inclusion of non-monotonic safety factor profiles is thought to improve plasma confinement, since this regime is capable to stabilize microinstabilities  (trapped electron modes) and magnetohydrodynamic (MHD) instabilities \cite{levinton1995,strait1995}. The non-monotonicity associated to the profile is responsible for new characteristic of the map, as the existence of twin islands and separatrix reconnection process, and it also  influences the types of bifurcation that can occur in the dynamics \cite{corso1998}.

{\color{black}The dynamics of plasma configuration with non-monotonic safety factor profiles can be studied by nontwist systems where the twist condition ${\partial x_{n_1}}/{\partial y_n} \ne0$ (also called non-degeneracy condition) is locally violated.} Nontwist maps violate the twist condition, which provides a monotonic relation between the momentum and the velocity in the phase space \cite{morrison2000}. The  behavior of nontwist dynamics systems can be explained by the Standard Nontwist Map (SNM), a paradigmatic two dimensional map proposed by Del-Castillo Negrete and Morrison \cite{morrison1993}. Some properties of plasmas in a non-monotonic scenario can be explained by the SNM analysis, but the map does not present a direct relation between its parameters and the physical quantities of tokamaks experiments. 

Others nontwist maps were developed in order to describe some properties of the magnetic field lines in confined plasmas with non-monotonic profile, as the Nontwist Ullmann map \cite{viana1992,ullmann2000} and the Revtokamap \cite{balescu1998}. The Ullmann map is a symplectic two dimensional map designated to describe the magnetic field lines in a tokamak with ergodic limiter. The map admits general safety factor profiles and presents an arbitrary parameter responsible for the toroidal correction and with agreement with experimentally observed profile \cite{ullmann2000}. Since the map admits arbitrary profiles, the inclusion of a non-monotonic profile is investigated in Refs. \cite{portela2007,viana2021,portela2008,caldas2012-2,bartoloni2016}. The nontwist version of the Ullmann map is equivalent to the standard nontwist map with the addition of a term, related to the toroidal correction, around the non-monotonic region \cite{portela2007}. The Revtokamap \cite{balescu1998}, on other hand, is composed by the inclusion of a non-monotonic profile in the Tokamap. The Tokamap, proposed by Balescu and coauthors, is not directly derived from the study of magnetic field lines, though it can represent some aspects of the global dynamics of the field lines in tokamaks \cite{portela2008,balescu1998-1}.

Following the idea of including non-monotonic profiles in existent maps, we propose, in this paper, the inclusion of a non-monotonic safety factor profile in the original Martin-Taylor (OMT) map. In order to do this, we generalize this map such that it admits an arbitrary safety factor profile. With a non-monotonic profile, we obtain a nontwist version of the OMT map, the non-monotonic extended Martin-Taylor (EMT) map. In order to understand the role of non-monotonicity in the map, we analyze and compare the map with monotonic and non-monotonic profiles by investigating the resonances produced by these systems. The EMT map with non-monotonic safety profile exhibits typical properties of nontwist systems, like twin island chains, shearless curves and separatrix reconnection \cite{morrison2000, portela2007, viana2021}. Using the pendulum approximation, we obtain expressions for the position and the width of resonant islands. We find that, for the nontwist case, this approximation is acceptable only for a limited range of the perturbation. Thanks to the nontwist characteristic of the map, the shearless transport barrier is robust and can be broken only for relatively large perturbation strength, in contrast with the twist case, for which the invariant curves are broken accordingly to their winding numbers.

This paper is organized as follows: we present the generalized version of the Martin-Taylor map and its nontwist version in Section 2. In Section 3, we study the resonances present in the system by the pendulum approximation applied to both twist and nontwist maps. Since the inclusion of non-monotonic profile causes the emergence of the shearless barrier, we study the effect of the perturbation parameters in the shearless curve in Section 4. Our conclusions are presented in Section 5.

\section{Extended Martin-Taylor map}
The effect of an ergodic magnetic limiter (EML) on the structure of the magnetic field lines in Tokamaks was studied by Martin and Taylor in their seminal paper of 1984 \cite{martin1984}. In order to describe the effect of the external perturbation due to the limiter on the magnetic surfaces, they proposed a two dimensional area preserving map composed by two maps where one describes the effect of the limiter and the other represent the influence of the shear in the field lines \cite{portela2003,martin1984}.
	
For the deduction of the map, Martin and Taylor considered a tokamak with large aspect ratio, such that the toroidal curvature can be neglected and the tokamak can be approximated to a periodic cylinder \cite{portela2003}. Furthermore, since the purpose of the map is to describe the behavior near the tokamak wall, the poloidal curvature can also be neglected such that it is possible to use a rectangular coordinate system \cite{portela2003}.  In such geometry, the coordinate $x=b\theta$ stands for the rectified arc along the tokamak wall, where $b$ is the minor radius of the tokamak, $\theta$ represents the poloidal angle, and $y=b-r$ is the radial distance to the tokamak wall at $y=0$. They define discretized variables $(x_n,y_n)$ as the values of the field line coordinates at a fixed Poincaré surface of section. With these approximations, the original Martin-Taylor (OMT) map is defined by the composition $M=T_1\circ T_2$, where \cite{portela2003}
\begin{widetext}
\begin{eqnarray}
	T_2:\begin{cases}
		x_n^*=x_n-\dfrac{bp}{m} \exp\left(\dfrac{-m y_n}{b}\right) \cos\left(\dfrac{m x_n}{b}\right),\\~\\
		y_n^*=y_n+\dfrac{b}{m}\ln\left\{ \cos \left[\dfrac{m x_n}{b}- p \exp\left(\dfrac{-m y_n}{b}\right) \cos\left(\dfrac{m x_n}{b}\right) \right] \right\}
  -\dfrac{b}{m}\ln\left[\cos\left(\dfrac{m x_n}{b}\right)\right],
		\end{cases}
	\label{T2}
\end{eqnarray}
\end{widetext}
represents the action of the ergodic limiter, \textit{i.e.}, the magnetic field line enters the limiter at the point with coordinates $(x_n,y_n)$ and emerges at a point with $(x_n^*,y_n^*)$ \cite{martin1984}. The parameter $m$ is the number of toroidal pairs of coils segments that conduct a current $I$, and $p$ is the strength of the limiter action, and it is proportional to the limiter current \cite{portela2003,martin1984}. {\color{black} The map $T_1$ is defined by,
\begin{eqnarray}
T_1:\begin{cases}
x_{n+1}=x^*_n+\alpha+sy^*_n,\\
y_{n+1}=y_n^*
\label{T1}
\end{cases}
\end{eqnarray}
where $s$ is the shear parameter which indicates how the safety factor varies with the radius and $\alpha$ is defined as $\alpha=2\pi b/q_b$, with $q_b$ being the safety factor at the tokamak wall \cite{portela2003, portela2008}. The mapping $T_1$ describes the behavior of the field lines under the influence of the shear only, which results in a displacement in $x$ on a constant value of $y$ \cite{portela2003,martin1984}.} 
	
The map (\ref{T2})-(\ref{T1}) can be generalized in a way such that the safety factor profile $q(y)$ appears explicitly, which we call extended Martin-Taylor (EMT) map, 
\begin{eqnarray}
	\begin{aligned}
		x_{n+1}=& x_n-\dfrac{p}{m} e^{-m y_n} \cos(m x_n) + \dfrac{2 \pi}{q(y_{n+1})},\\
		y_{n+1}=& y_n+\dfrac{1}{m}\\
    &\times\ln\left\{ \dfrac{\cos[m x_n - p e^{-m y_n} \cos(m x_n)]}{\cos(m x_n)}\right\},
	\end{aligned}
	\label{mt}
\end{eqnarray}
where $x$ and $y$ have been normalized with respect to the Tokamak minor radius $b$. It can be verified that the map (\ref{mt}) is area-preserving for any profile $q(y)$. {\color{black} Therefore, the modification of the Martin-Taylor map allows us to include different types of safety factor profile as non-linear, monotonic and non-monotonic profiles. The original Martin-Taylor map used a linear approximation for the safety factor profile which cannot describe some aspects of the magnetic field in the inner plasma column, for example non-monotonic profiles. The interest in the extended Martin-Taylor map is that it can be adapted to a large number of possible safety factor profiles, providing a simple map for field lines exhibiting chaotic behavior in a variety of situations.} For the OMT map (\ref{T2})-(\ref{T1}), the safety factor profile is a linear approximation near the tokamak wall, i.e.,
\begin{eqnarray}
	q(y)=\dfrac{2\pi b}{(\alpha+sy)}=\dfrac{2\pi q_b}{2\pi+s q_b~ y}.
	\label{qmonot}
\end{eqnarray}
{\color{black} with $y$ normalized with minor radius $b$. }

Since $q(y)$ is arbitrary, we can use a non-monotonic profile which matches some plasma discharges that occur in tokamak experiments \cite{portela2007,oda1995,caldas2012}. The non-monotonic profile is chosen by considering a current density with a non-monotonic behavior, a central hole and a peak outside the center \cite{oda1995,stix1976,roberto2004,martins2011}, 
\begin{eqnarray}
	\mathbf{j}=j_0\left(1+\beta' \dfrac{r^2}{a^2}\right)\left(1-\dfrac{r^2}{a^2}\right)^\mu \hat{\mathbf{z}},
	\label{j}
\end{eqnarray}
where $a$ is the plasma radius, which is usually slightly less than the minor radius $b$, and $j_0$ is proportional to the total plasma current. The parameter $\beta'$ is defined as $\beta'=[\beta(\mu+1)]/(\beta+\mu+1)$, and $\mu$ and $\beta$ are chosen to fit the profiles observed experimentally \cite{portela2007,caldas2012}. For the current density (\ref{j}), the resulting safety factor profile is written, in the variables $x$ and $y$, as \cite{portela2007,caldas2012},
\begin{widetext}
\begin{eqnarray}
	q(y)=q_a \dfrac{(1-y)^2}{a^2} \left[1-\left(1+\beta'\dfrac{(1-y)^2}{a^2}\right)\left(1-\dfrac{(1-y)^2}{a^2}\right)^{\mu+1}\Theta(a+y-1)\right]^{-1}
	\label{q-profile}
\end{eqnarray}
\end{widetext}
where $\Theta$ is the unit step function and $q_a=q(r=a)$ is the value of the safety factor at the edge of the plasma and it is also chosen to fit experimental results \cite{portela2007}. {\color{black} If $\beta'=0$ for the safety factor profile (6), the profile is monotonic in the range $y\in[0,1]$ and approaches the behavior of the safe profile for the OMT.} The shear parameter $s$ is given by
\begin{eqnarray}
	s=-\dfrac{2\pi b}{q^2} \dfrac{dq}{dy}.
	\label{shear}
\end{eqnarray}

\begin{figure*}
	\centering
	\includegraphics[width=0.9\textwidth]{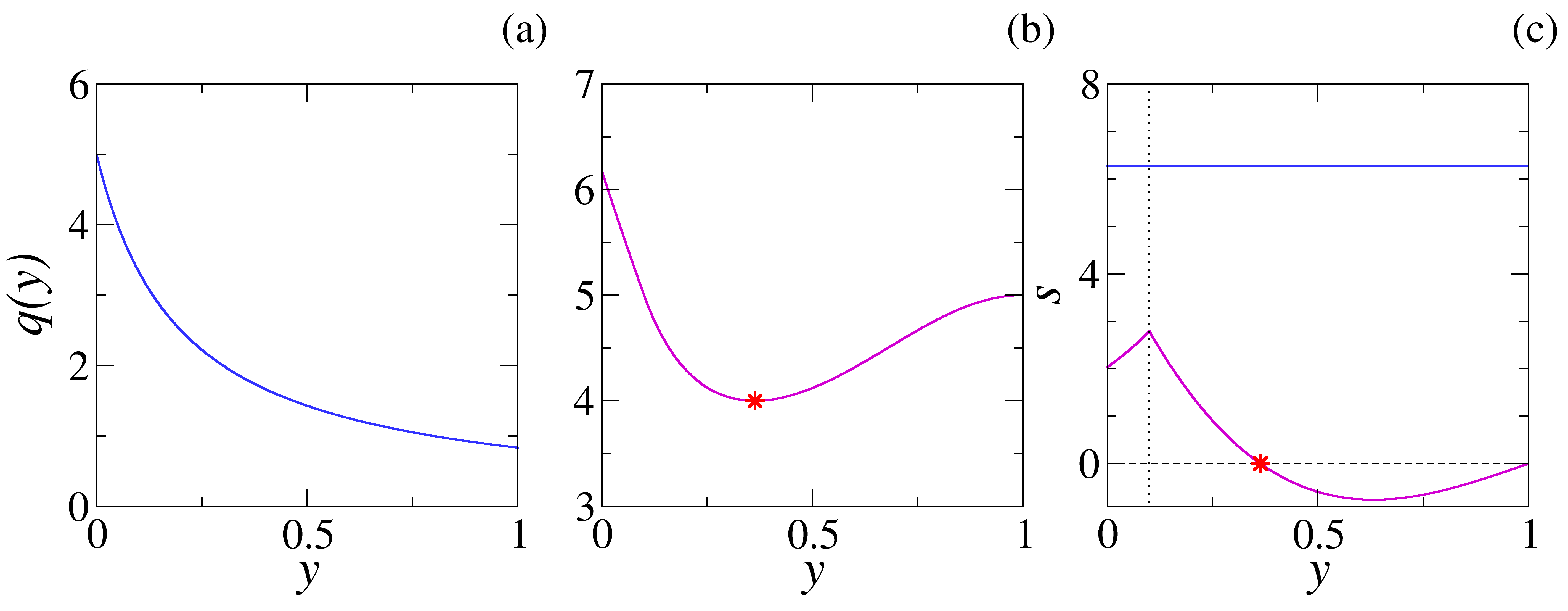}
	\caption{Examples of safety factor profiles $q(y)$. The monotonic profile (\ref{qmonot}) of the OMT map is exhibited in (a) and the parameters are $s=2\pi$ and $q_b=5$. For the non-monotonic profile  of the EMT map in (b), we choose the parameters $q_a=5$, $\beta=2$, $\mu=1$ and $a=0.9$. The shear respective to the monotonic and non-monotonic profile are shown in panel (c) by the constant (blue) value $s=2\pi$ and by the curve, respectively. The dotted line in (c) indicates the edge of the plasma while the dashed line highlights the point in which the shear is null. (Color available online)}
	\label{fig1}
\end{figure*}

In the OMT map, the shear parameter takes a constant value, which we take as $s=2\pi$ \cite{portela2003,martin1984}. For the non-monotonic profile (\ref{q-profile}) of the EMT map, the respective shear is\cite{portela2003,martin1984},
\begin{widetext}
\begin{eqnarray}
	\begin{aligned}
		s=\dfrac{4\pi a^2}{q_a (1-y)^3}&\bigg\{1- \\
		&- \left.\left(1-\dfrac{(1-y)^2}{a^2}\right)^\mu \left[1+\dfrac{(1-y)^2}{a^2}\left(\mu+\beta'(1+\mu)\dfrac{(1-y)^2}{a^2}\right)\right]\Theta(a+y-1)\right\}. 
	\end{aligned}
\end{eqnarray}
\end{widetext}
The safety factor radial profiles for both the OMT map, Eq. (\ref{qmonot}), and the EMT map, Eq. (\ref{q-profile}), are shown in Figure \ref{fig1}, as well as the respective values of the shear parameter. For the OMT map, $s$ is constant and indicates a monotonic increase from the center $(y=1)$ to the edge $(y=0)$ of the plasma [Figure \ref{fig1} (a)]. For the non-monotonic profile of EMT map showed in Figure \ref{fig1} (b), we observe an extremum at $y\approx 0.36$, where the profile has a minimum value. {\color{black} There is a radial range in which $q$ is not single-valued}, what does not occur for the monotonic case.
	
From the shear of the non-monotonic profile, plotted as the pink curve in Figure \ref{fig1} (c), we draw attention to two points: the sharp turn at $y=0.1$, and the null value in $y\approx0.36$. The sharp point occurs at the edge of the plasma $r=a=0.9$ ($y=0.1$), distinguishing the regions inside and outside the plasma. The shear vanishes at $y\approx 0.36$, highlighted by the dashed line. This point corresponds to the minimum value of $q$ in Figure \ref{fig1} (b) and, consequently, to the point where the twist condition, ${\partial x_{n+1}}/{\partial y_n} \ne 0 $, is violated by the system \cite{morrison1993}. In the dynamical system, this point belong to the shearless curve, a typical solution of degenerate systems, characterized by having a extreme value of the winding number \cite{shinohara1997}. {\color{black} The winding number $\omega$ associated to a solution with initial condition $(x_0,y_0)$ is defined by the limit \cite{del1996},
\begin{eqnarray}
	\omega=\lim_{n\to \infty} \dfrac{x_n-x_0}{n},
 \label{wn}
\end{eqnarray}
that converges only if the solution is regular (quasi-periodic or periodic).}
	
\begin{figure}[!h]
	\centering
	\includegraphics[width=0.45\textwidth]{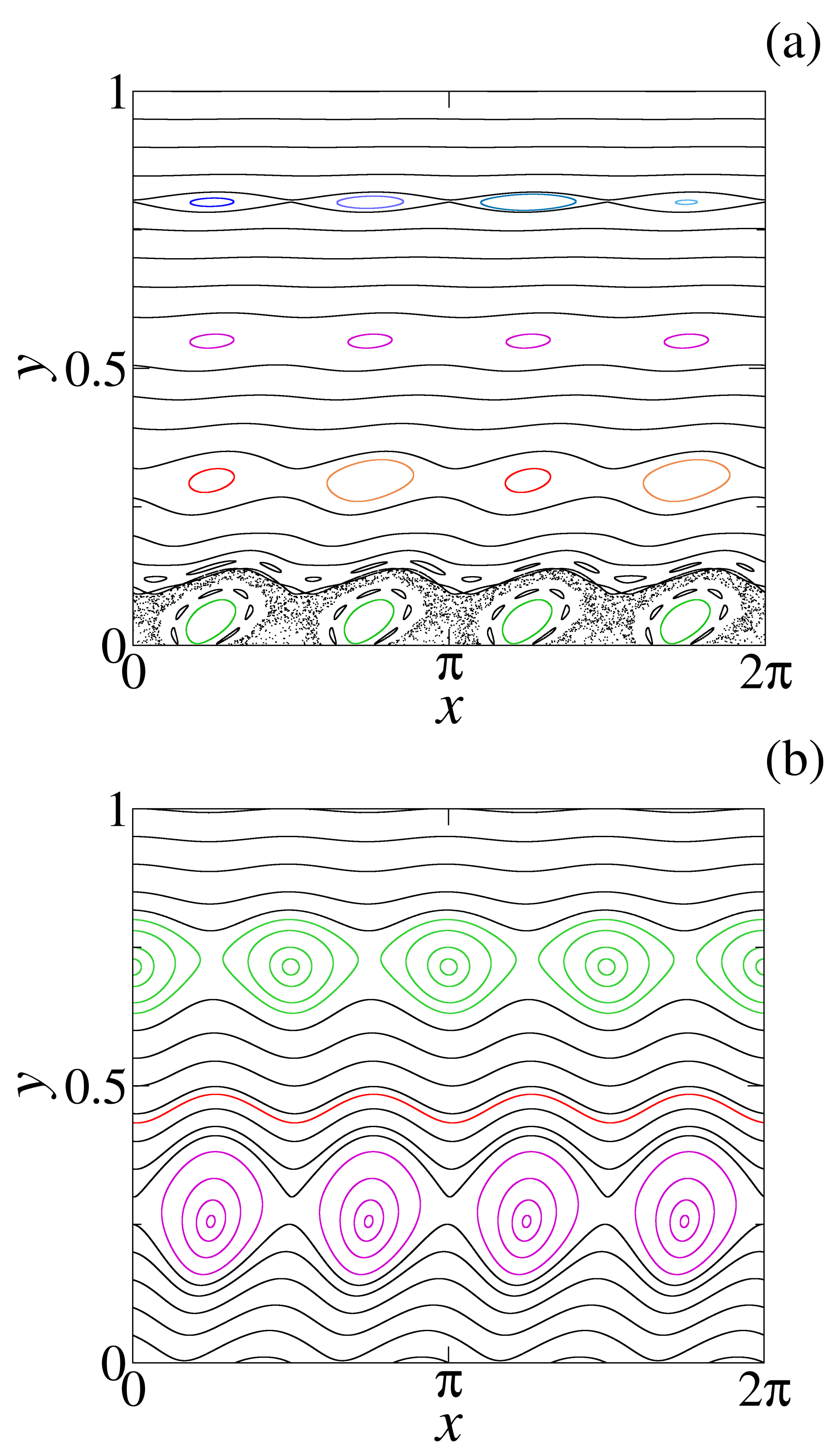}
	\caption{Comparison between the phase spaces for the (a) OMT and (b) EMT map. For both spaces, $m=4$, $p=0.2$ and $q_a=q_b=4.5$. For the extended map, the plasma radius is $a=0.9$ and the shearless curve is indicated by the curve in red in (b).}
	\label{fig2}
\end{figure}

In order to examine the influence of the safety factor profiles on the EMT map, {\color{black}  we construct the phase space by plotting some solutions of} OMT map (\ref{mt}) and EMT map with non-monotonic profile (\ref{q-profile}). The phase spaces are shown in Figure \ref{fig2}.
	
The phase space for the OMT map [Figure \ref{fig2} (a)] depicts a typical situation for conservative near-integrable systems. We observe the coexistence of periodic and quasi-periodic solution, represented by the solid curves, with a chaotic area-filling orbit represented by dispersed points at the lower region of the phase space, near the edge of the tokamak. Since $m=4$, we observe multiple chains of four islands centered at different values of $y$. However, not all the chains belong to the same resonance: the four {\color{black} lower} (green) islands  are generated by the same initial condition, as the islands {\color{black} at the center of the phase space (center aroud $y=0.55$)}. For the islands in shade of red around $y\approx 0.3$, however one initial condition generate two islands, and therefore we observe islands with two different sizes. A similar scenario is observed for the islands around $y\approx 0.8$. These islands are not connected, since they correspond to four different solutions.
	
\begin{figure}[!h]
	\centering
	\includegraphics[width=0.45\textwidth]{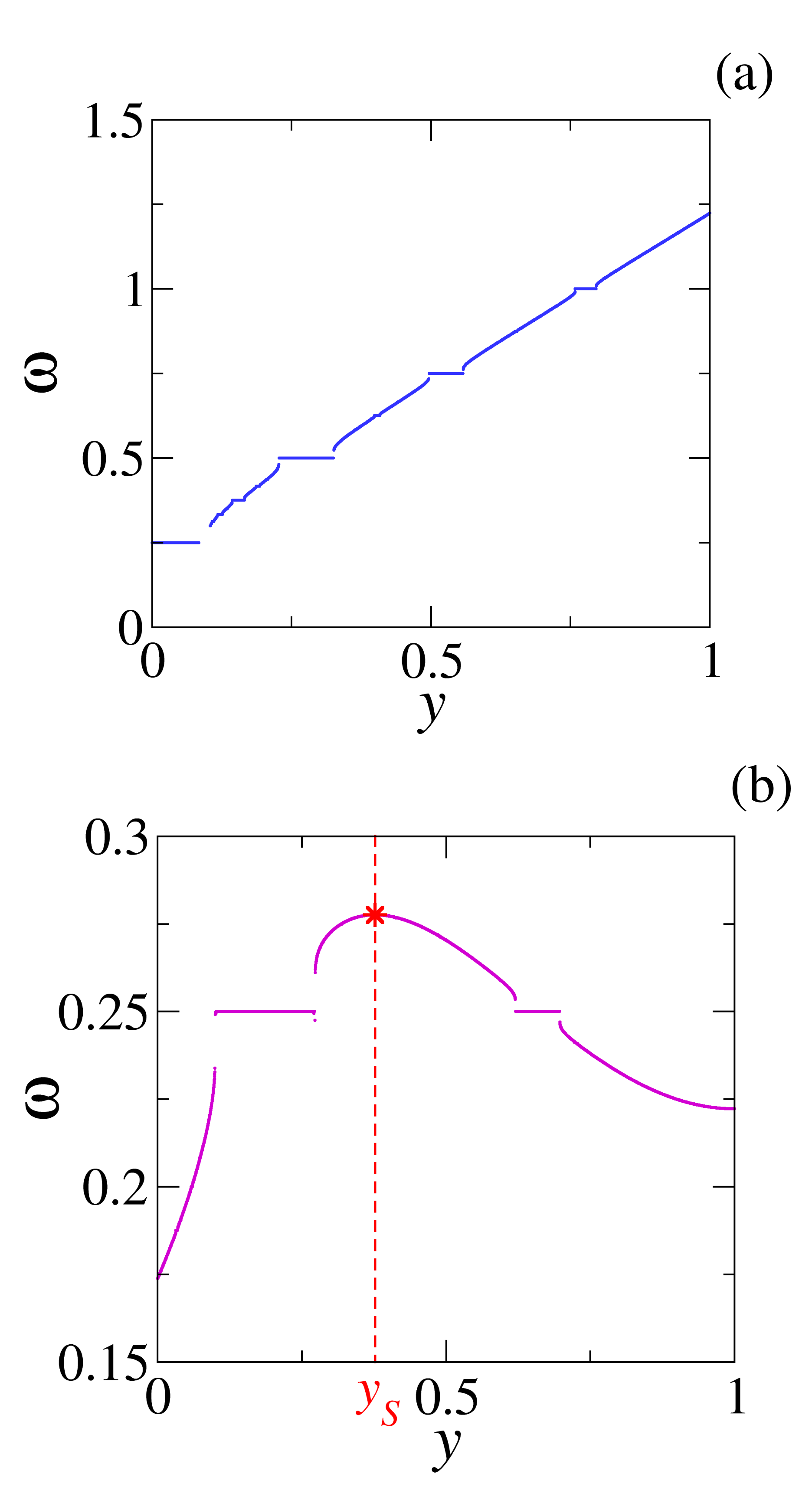}
	\caption{Winding number profile for the (a) OMT map and the (b) EMT map. The chosen lines for the computation is $x_0=0.8$ and $x=1.1$, respectively. These lines are indicated in Figure \ref{fig2} by the dashed lines. The point $y_S$ indicates the position of the shearless curve in the profile.}
	\label{fig_wn}
\end{figure}

For the EMT map [Figure \ref{fig2} (b)] chaotic behavior is not present in the phase space, indicating that the perturbation is not strong enough to break up the regular solutions. Additionally, we observe rotational circles represented by the black curves, and two chains of four islands each. For both chains, all the islands belong to the same solution. In red, we have the shearless curve in which the winding number assumes its maximum value. In order to highlight the behavior of the winding number for both maps, we compute profile of $\omega$ for a line $x_0$ in the phases spaces of Figure \ref{fig2} and present the respective profiles in Figure \ref{fig_wn}. For each point $(x=x_0,y)$, we computed the limit (\ref{wn}), considering as final iteration time $n_F=10^6$.

In Figure \ref{fig_wn} (a), we observe the winding number profile for the phase space of the OMT map presented in Figure \ref{fig2} (a). The value of $\omega$ increases monotonically as $y$ varies from $y=0$ to $y=1$. The are some plateaus indicating the existence of islands in the phase space. Each plateau is related to the colored islands in the phase space of Figure \ref{fig2} (a). 

For the EMT map, the winding number profile is presented in Figure \ref{fig_wn} (b). We observe a non-monotonic scenario, in which $\omega$ takes on an extreme value at $y_S$, indicated by the red symbol. In this case, the point of extreme is a maximum. Due to the non-monotonic behavior of the winding number, at least two distinct solutions present the same $\omega$.  A solution below the shearless curve ($y<y_S$)  will have the same value of $\omega$ than a solution above the curve ($y>y_S$). This is what happens for the two islands chains in Figure \ref{fig2} (b). They are at different ``sides" of the shearless curve and present the same winding number, indicated by the plateaus in Figure \ref{fig_wn} at $\omega=0.25$. For this reason, they can be called ``twin" islands.
	
Since the map is not symmetric, the twin islands are different. These scenario were observed for other asymmetric nontwist maps \cite{mugnaine2020}. For the standard nontwist map, del-Castillo-Negrete, Greene and Morrison remarked that, when the period of the twin islands is even, the elliptic and hyperbolic points of each chain are aligned with each other in a way that, if the chains collide, a hyperbolic-hyperbolic collision happens \cite{del1996}. This is not observed in the EMT map. Observing the islands in Figure \ref{fig2} (b), if they eventually collide, a hyperbolic-elliptic collision would happen, the same collision that occurs when the islands presents odd period in the SNM. 
	
\begin{figure*}
	\centering
	\includegraphics[width=0.85\textwidth]{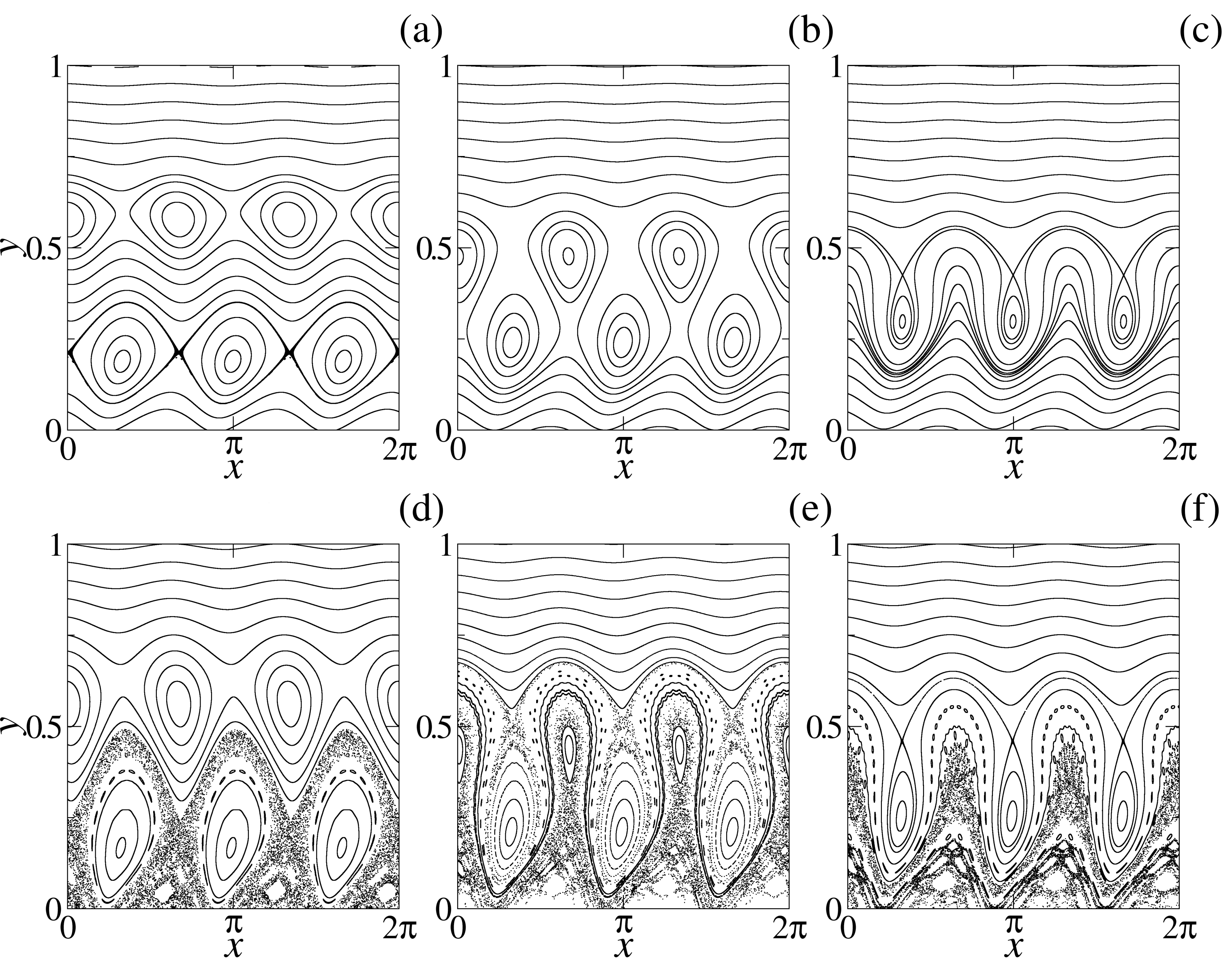}
	\caption{The effect of the parameters $p$ and $q_a$ in the phase space for the extended Martin-Taylor map. For the first line, we have $p=0.15$ while in the second line, $p=0.4$. The first, second and third columns are phase spaces for $q_a=3.5$, $q_a=3.65$ and $q_a=3.75$, respectively. For all phase spaces, $a=0.9$ and $m=3$.}
	\label{fig3}
\end{figure*}

{\color{black} The EMT map presents four control parameters: 1) the perturbation amplitude $p$ that, as discussed for the OMT map in the previous section, represents the limiter action, 2) the plasma radius $a$, 3) the safety factor value at the plasma edge $q_a$ and 4) the number of pairs of toroidal oriented segments $m$ \cite{portela2003}}. In Figure \ref{fig3}, we consider the influence of $q_a$ and $p$ in the phase space for fixed $a=0.9$ and $m=3$.

The phase spaces in Figure \ref{fig3} are computed for two values of perturbation: $p=0.15$, representing a weak perturbation, and $p=0.4$, a strong perturbation. The small perturbation scenario is shown in the upper panels, (a), (b) and (c), while the lower panels, (d), (e) and (f), portray the scenario for the strong perturbation. For the first, second and third column, we have $q_a=3.5$, $q_a=3.65$ and $q_a=3.75$, respectively.

Observing the structures in the phase spaces for each $p$ following the increase of $q_a$, we notice that the $q_a$ is related to the separatrix reconnection scenarios, similar to the scenarios described for the SNM \cite{del1996}. As the parameter increases and the two chains of islands approach, a hyperbolic-elliptic collision occurs. The upper (lower) chain of islands that was above (below) the shearless curve changes its side after the reconnection and the lower chain of island is destroyed for the largest value of $q_a$ considered. This sequence occurs for both values of $p$, with the difference that for the larger value of $p$ we have a chaotic sea surrounding the lower chain. 
	
With the results presented by the phase spaces in Figure \ref{fig3}, we can conclude that the increase of the perturbation parameter $p$ implies the eventual emergence of chaotic behavior in the system, while the safety factor $q_a$ is related to the scenarios of separatrix collision/reconnection. This scenario was observed in Poincaré maps of numerically integrated magnetic field line trajectories obtained for MHD equilibrium plasmas in tokamaks \cite{corso1997}.  In our simulation, we observe that a variation in the value of $a$ changes the value of $y$ in the position of the islands. Then, we choose $a=0.9$ since the two twin islands and the reconnection process, showed in Figure \ref{fig3}, occurs for the range of the parameters $p$ and $q_a$ chosen in this study. The parameter $m$ is related to the resonances and island periods.

\section{Hamiltonian and pendulum approximation}
The OMT map is area-preserving, as much as it is not directly derived from a Hamiltonian function. Portela and coauthors treat the map as a near-integrable system and were able to write the map as a perturbed twist mapping \cite{portela2003}. With this approach, they were able to find the fixed points for small values of $p$, study their stability and estimate the limiter strength threshold for the transition to global chaos as previously analyzed by Martin and Taylor \cite{portela2003,martin1984}.

The Hamiltonian function $H=H(x,y,n)$ for the EMT map (\ref{mt}) is, for small values of $p$ (see the Appendix for details),
	
\begin{eqnarray}
\begin{aligned}
	H= 2\pi\int \dfrac{dy}{q(y)}\dfrac{p}{m^2} e^{-my}\cos(mx) \delta(n),
	\label{hamilt}
\end{aligned}
\end{eqnarray}
{\color{black} where the periodic delta function is included in order to transform difference equations into differential equations, as showed in Ref. \cite{lichtenberg} (p. 171).}

The Hamiltonian function (\ref{hamilt}) can be written in the form $H=H_0(y)+\epsilon H_1(x,y,n)$, as it is usual for Hamiltonian near-integrable systems constructed by the addition of a perturbation $H_1$ with magnitude $\epsilon$ to the integrable systems $H_0$ \cite{lichtenberg,ott}. Using

\begin{eqnarray}
\begin{aligned}
    \delta(n)&=\sum_{k=-\infty}^{\infty} \delta(n-k)\\
    &=1+2\sum_{k=1}^{\infty} \cos(2\pi k n),
    \end{aligned}
\end{eqnarray}
the perturbation in (\ref{hamilt}) can be rewritten as,
\begin{eqnarray}
\begin{aligned}
	H_1(x,y,n)=\dfrac{p}{m^2}e^{-my} & \sum_{k=0}^{\infty} \left[\cos(mx+2\pi n k)\right.\\~\\~\\ & \left.+\cos(mx-2\pi n k)\right].
	\label{h1}
 \end{aligned}
\end{eqnarray}
The resonance in the system happens when $mx-2\pi n k$ assumes a constant value. This condition leads to the following results:
\begin{enumerate}
	\item The terms $\cos(mx+2\pi n k)$ vary rapidly with time and vanish after an average performed over $n$. The Hamiltonian (\ref{h1}), near a resonance, reduces to,
	\begin{eqnarray}
	H_1=\dfrac{p}{m^2} e^{-my} \sum_{k=0}^{\infty} \cos(mx-2\pi n k).
	\end{eqnarray} 
	\item For a resonance $k$, the time evolution of $x$ is governed by,
	\begin{eqnarray}
		\dfrac{dx}{dn}=\dfrac{2\pi k }{m}.
	\end{eqnarray} 
	\item The safety factor profile for the unperturbed system, at the resonance, is a rational number,
	\begin{eqnarray}
		q(y^*)=\dfrac{m}{k},
	\end{eqnarray}
where $y^*$ is the location of the resonance in the phase space.
\end{enumerate}

Next to the resonance, we can consider $y=y^*+\Delta y$, where $|\Delta y| \ll |y^*|$, and expand the Hamiltonian function around the resonance point $y^*$. The resulting Hamiltonian function is,

\begin{widetext}
\begin{eqnarray}
\begin{aligned}
	\Delta H = \dfrac{2\pi K}{m} \Delta y +2\pi &\left( \dfrac{d}{d y}  \left. \dfrac{1}{q(y)}\right)\right|_{y^*}  \dfrac{(\Delta y)^2}{2}+\\
 &+\dfrac{p~ e^{-my^*}~ \cos(mx-2\pi nk)}{m^2} \left(1-m\Delta y +\dfrac{m^2}{2} (\Delta y)^2 + ...\right).
 \label{expan}
 \end{aligned}
\end{eqnarray}
\end{widetext}

We perform a canonical transformation $(\Delta y, x,n) \to (I, \theta)$ with the generating function,
\begin{eqnarray}
	F_2(I,x,n)=(mx-2 \pi n k ) I.
	\label{f2}
\end{eqnarray}
In a first approximation, considering only the first term in expansion in parenthesis, the result is the pendulum Hamiltonian,
\begin{widetext}
\begin{eqnarray}
\begin{aligned}
	\mathcal{H}= 2\pi m^2 \left(\dfrac{d}{d y} \left.\dfrac{1}{q(y)}\right)\right|_{y^*} \dfrac{I^2}{2}+\dfrac{p~ e^{-my^*}}{m^2} ~ \cos(\theta) = G \dfrac{I^2}{2} + F \cos(\theta).
	\label{Happrox}
 \end{aligned}
\end{eqnarray}
\end{widetext}

Using this result, we can relate the results obtained with the pendulum approximation (\ref{Happrox}) with the actual results for the EMT map (\ref{mt}), for small values of the perturbation amplitude $p$. The half-width of an island, centered at $y=y^*$, is computed by \cite{lichtenberg},
\begin{eqnarray}
	I_{max}= 2\sqrt{\left|\dfrac{F}{G}\right|},
 \label{imax}
\end{eqnarray}
while the frequency near the resonance is $\omega^*=\sqrt{F G}$.

In this section, we aim to study the validity of pendulum approximation for the OMT and EMT map, focusing on the first resonance of the system. We are interested in the maximum half-width of the islands with $k=1$, for different values of $m$, and in the range of $p$ where the approximation is valid. We first study the approximation for the OMT map and compare with the results for the EMT map, in order to identify the effect of non-monotonicity in the resonance structure of the system. 

For the purpose of relating the value of $p$ with the parameters related to the tokamak, we take the definition of $p$ \cite{portela2003}, 
\begin{eqnarray}
    p=\dfrac{\mu_0 \ell I_L}{B_0 b^2 \pi},
    \label{eqp}
\end{eqnarray}
where $\ell$ is the finite length of the EML, $B_0$ is the toroidal equilibrium field, $I_L$ represents the magnitude of the limiter current. The safety factor at the edge of the plasma is computed by \cite{wesson},
\begin{eqnarray}
    q_a=\dfrac{2\pi a^2 B_0}{\mu_0 I_P R_0},
    \label{eqq}
\end{eqnarray}
where $I_P$ is the total current of the plasma and $R_0$ is the major radius of the tokamak. Combining the equation (\ref{eqp}) and (\ref{eqq}), we obtain the relation between the strength $p$ and the ratio of the currents $I_L / I_P$,
\begin{eqnarray}
    p=\dfrac{2m^2 \xi a^2}{q_a}  \dfrac{I_L}{I_P},
\end{eqnarray}
where $\xi=\ell/R_0$ and $a/b \to a$. In the present paper, we fixed $\xi=0.163$ according to the TCABR parameters $\ell=0.1$ m and $R_0=0.615$ \cite{pires2005}.

\subsection{Original Martin-Taylor Map}

From Eq. (\ref{Happrox}) with the safety factor profile (\ref{qmonot}), we are able to compute the Hamiltonian function for the OMT map near a resonance, where the Hamiltonian is,
\begin{eqnarray}
\begin{aligned}
    H_{OMT}=\dfrac{s m^2}{2} &I^2 +\\
    +&\dfrac{p}{m} \exp\left[-\dfrac{2\pi}{s}\left(k-\dfrac{m}{q_b}\right)\right]\cos \theta,
    \label{HamiltMT}
\end{aligned}
\end{eqnarray}
and the resonance $k/m$ occurs at,
\begin{eqnarray}
    y^*=\dfrac{2\pi (kq_b-m)}{s m q_b}.
    \label{resonMT}
\end{eqnarray}
For the resonance to occur in the physical domain $0\le y\le 1$, the parameters $k$, $m$ and $q_b$ must satisfy the relation,
\begin{eqnarray}
    1\le \dfrac{k q_b}{m} \le 1 + \dfrac{s q_b}{2\pi}.
    \label{condMT}
\end{eqnarray}

From the pendulum approximation, the half-width of the island around the resonance for the OMT map, given by (\ref{imax}) is,
\begin{eqnarray}
\begin{aligned}
I_{max}&=\frac{2}{m^2}\left\{\dfrac{2 p}{s}\exp\left[-\dfrac{2\pi}{s}\left(k-\dfrac{m}{q_b}\right)\right]\right\}^{1/2}\\
I_{max}&\sim p^{1/2}.
\end{aligned}
\label{ImaxTeorico}
\end{eqnarray}

\begin{figure*}
	\centering
	\includegraphics[width=0.9\textwidth]{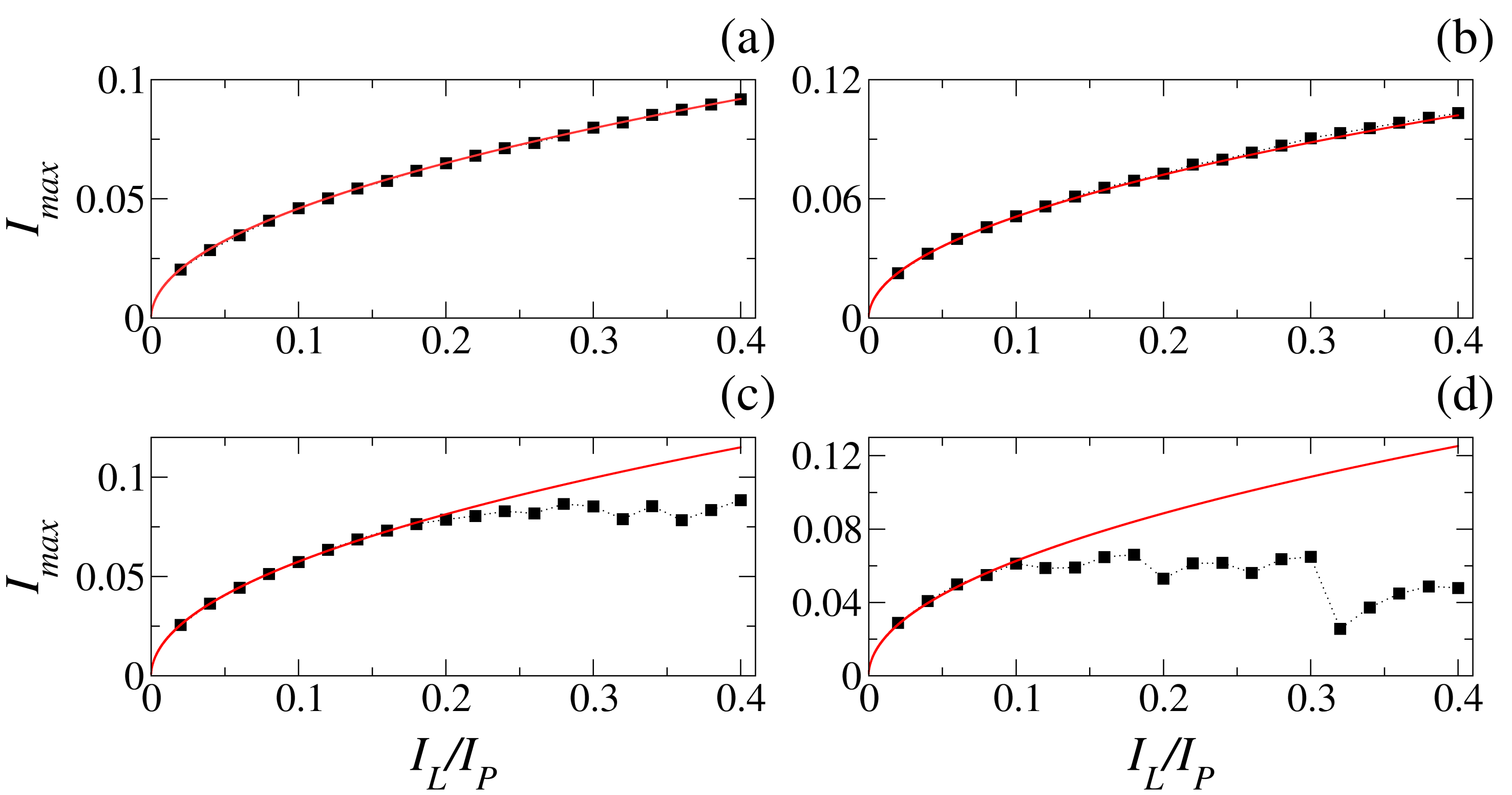}
	\caption{For the OMT map, comparison between the half-width of the islands related to the resonance $k=1$ obtained numerically (black squares) and by the equation (\ref{ImaxTeorico}) from pendulum approximation ({\color{black} continuous} curve), in relation to the ratio between the limiter current ($I_L$) and the total plasma current ($I_P$), for $q_b=4.5$. The half-widths were computed for four values of $m$: (a) $m=1$, (b) $m=2$, (c) $m=3$ and (d) $m=4$.}
	\label{fig4}
\end{figure*}

In order to verify the validity of Eq. (\ref{ImaxTeorico}), we estimate numerically the half-width of the largest island in the phase space resultant of the map iteration. The half-width is calculated by the difference ${(y_{max}-y_{min})}/{2}$ where $y_{max}$ and $y_{min}$ are the maximum and the minimum value of $y$ assumed by the largest $k/m$-island of the phase space, respectively. The results of the half-width calculated by (\ref{ImaxTeorico}) and numerically are shown in Figure \ref{fig4}.

In Figure \ref{fig4}, the {\color{black} continuous} curves indicate the $I_{max}$ computed by (\ref{ImaxTeorico}) while the black squares are the half-width calculated by the analysis of the islands in the phase space. Comparing both $I_{max}$, we observe that for $m=1$ and $m=2$, Figures \ref{fig4} (a) and (b), the numerical results agree with the theoretical estimate. The half-width increases with $(I_L/I_P)^{1/2}$, at least for the range we analyzed.  A different scenario is observed for $m=3$ and $m=4$. For $m=3$, Figure \ref{fig4} (c), we observe that the numerical values increase with $(I_L/I_L)^{1/2}$ for $I_L/I_P <0.2$. For larger values of the ratio, the value of $I_{max}$ varies somewhat close to $I_{max}=0.083$, indicating that if the limiter current is larger than $20\%$ of the plasma current, the pendulum approximation is no longer valid. 

The scenario for $m=4$ is shown in Figure \ref{fig4} (d). We observe an agreement between the theoretical and numerical solution up to $I_L/I_P \approx 0.1$. For larger values, the numerical value of $I_{max}$ fluctuates about $I_{max} \approx 0.06$ and then decreases for $I_L/I_P > 0.3$. Computing the phase spaces for the system in the range where the decrease happens, there is an enlargement of the chaotic sea around the islands and, consequently, their shrinkage. Calculating the position of the resonance, given by (\ref{resonMT}), for $m=3$, $m=4$ and the parameters used for the results in Figure \ref{fig4}, we obtain {\color{black}{$y^*_{m=3}=1/9$}} and {\color{black}{$y^*_{m=4}=1/36$}}, respectively, indicating that the islands are close to the wall ($y=0$). Once the chaotic solutions firstly emerge in this region, the islands for $m=3$ and $m=4$ suffer the effect of the onset of chaos for smaller values of $I_P/I_L$, compared to the magnetic islands for $m=1$ and $m=2$. Therefore, we conclude that the existence of the stable orbits around the resonances is not affected only by the strength of the perturbation, but mostly because of their radial position $y$ that is a consequence of the safety factor $q_b$ in the edge of the plasma.

\subsection{Extended Martin-Taylor Map}

Due to a more complicated function of the non-monotonic profile $q(y)$ of (\ref{q-profile}), the Hamiltonian function (\ref{Happrox}) for the EMT map is given in an implicit form. The term $G$ and $F$ of the pendulum approximation are given by,

\begin{eqnarray}
\begin{aligned}
    G=\left.\dfrac{2 \pi m^2 a^2}{q_a} \dfrac{d q^{-1} (y)}{dy}\right|_{y^*},~ F= \dfrac{p e^{-my^*}}{m^2},
\end{aligned}   
\end{eqnarray}
respectively, where $q(y)$ is given by the profile (\ref{q-profile}), and $y^*$ is the solution of the equation,
\begin{widetext}
\begin{eqnarray}
\begin{aligned}
    \dfrac{k q_a}{ma^2} (1-y^*)^2+1-\left(1+\beta'\dfrac{(1-y)^2}{a^2}\right)\left(1-\dfrac{(1-y)^2}{a^2}\right)^{\mu +1} \Theta(a+y^*-1)=0.
    \end{aligned}
\end{eqnarray}
\end{widetext}
Since $G$ and $y^*$ do not depend on $p$, the $I_{max}$ (by the pendulum approximation for the EMT map) also increases with $p^{1/2}$.

Considering $\mu=1$, we are able to find analytically the value of $y^*$ and, consequently, of $G$, as follows:
\begin{widetext}
\begin{eqnarray}
\begin{aligned}
    y^*&=1-a\left[\dfrac{2\beta'-1\pm\sqrt{1+4\beta'(1-kq_a/m)}}{2\beta'}\right]^{1/2},\\
    G&=\pm\dfrac{4\pi m^2}{a q_a}\sqrt{1+4\beta'\left(1-\dfrac{k q_a}{m}\right)}\left[\dfrac{2\beta'-1\pm\sqrt{1+4\beta'(1-kq_a/m)}}{2\beta'}\right]^{1/2}.
    \label{resonEMT}
\end{aligned}    
\end{eqnarray}
\end{widetext}

From the definition of $y^*$, in (\ref{resonEMT}), we observed two values of $y^*$ for each set of parameters, one for each chain of the twin islands, a scenario observed in Figure \ref{fig2} (b). The existence and the position of the islands depend on $q_a$. For this reason, once we want to analyze the half-width of the twin islands with $k=1$, we choose appropriate values of $q_a$, for which both chains exist. For the twin islands to be located in the physical domain $0\le y \le 1$, the resonance position $y^*$ belongs to this domain if,
\begin{eqnarray}
    \dfrac{3}{a^2}-\dfrac{1}{a^4}-1\le\dfrac{kq_a}{m}\le 1 +\dfrac{1}{4\beta'}.
    \label{condEMT}
\end{eqnarray}
We emphasize that $y^*$ defined by (\ref{resonMT}) and (\ref{resonEMT}) as the conditions (\ref{condMT}) and (\ref{condEMT}) were computed not considering the perturbation. Thus, the islands associated with the resonance can present different $y^*$ as well as they may already have been destroyed by the perturbation effect even if the conditions (\ref{condMT}) and (\ref{condEMT}) are satisfied. 

\begin{figure*}
	\centering
	\includegraphics[width=0.9\textwidth]{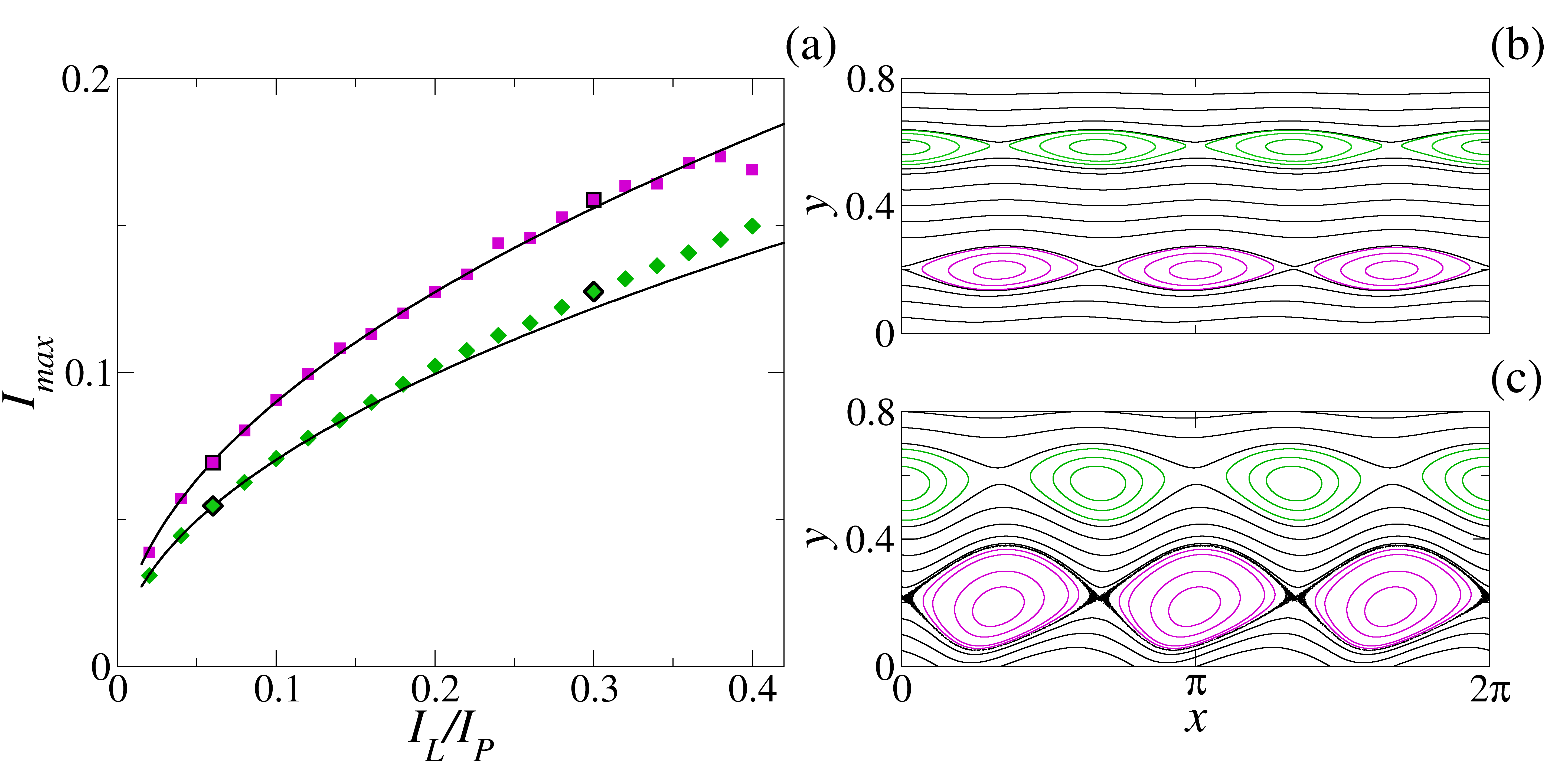}
	\caption{ Half-width of the islands for the EMT map. The values showed in (a) are associated to the islands from the lower (squares) and upper (diamond symbol) island chains for the extended Martin-Taylor map, related to the resonance $k=1$, in relation to the ratio $I_L/I_P$. The half-width are calculated numerically for $m=3$ and $q_a=3.5$. The black lines indicate the pendulum approximation (\ref{imax}). The phase spaces in (b) and (c) are related to the highlighted points in (a), $I_L/I_P=0.06$ and $I_L/I_P=0.3$, respectively.}
	\label{fig5}
\end{figure*}

As performed for the OMT map, we computed the half-width $I_{max}$ of the largest island in the phase space for different values of $I_L/I_P$ and verify if the scale law $p^{1/2}$ is observed. Once the system violates the twist condition, we computed the half-width for islands at both sides of the shearless curve. The islands in the upper and lower region of the phase space are named upper and lower islands and their half-widths are indicated by the green (diamond symbol) and magenta (squares) in Figure \ref{fig5}, respectively. The values of $I_{max}$ as a function of $I_L/I_P$, for the EMT map with $m=3$, are shown in Figure \ref{fig5}. The results observed for $m=1,~2$ and 4 are similar.

The half-widths presented in Figure \ref{fig5} (a) are calculated numerically, observing the phase space for each value of $I_L/I_P$. Once we choose appropriate values of $q_a$ in order to obtain the two chains of islands, the islands are not immersed in a large chaotic sea. In fact for most values of $(I_L/I_P)$ the chaotic behavior around the lower islands is restricted. Therefore, $I_{max}$ does not change significantly for higher values of $I_L/I_P$ as was observed for the OMT map for $m=3$ and $m=4$ in Figure \ref{fig4} (c) and (d).

Comparing the values obtained numerically with the pendulum approximation $I_{max} \propto ({I_L}/{I_P})^{1/2}$, represented by the black curves in Figure \ref{fig5} (a), we observe a disagreement between them for larger values of the ratio, $I_L/I_P > 0.2$. However, unlike it was observed in the case of the OMT map for $m=3$ and $m=4$, showed in Figures \ref{fig4} (c) and (d), the half-width of the islands for the EMT map does not decrease or fluctuates about a certain value. In fact, for the upper island chain, the half-width continues to increase, but does not follow the pendulum approximation relation. The same is identified for the lower island, we observe a general increase behavior, but the half-width does not increase with $ (I_L/I_P)^{1/2}$.

We choose two values of $I_L/I_P$ and construct the respective phase spaces. The ratio $I_L/I_P=0.06$ represents the case where the numerical value is close to that predicted by the pendulum approximation. The phase space is shown in Figure \ref{fig5} (b). For the scenario where the numerical and expected value of $I_{max}$ are comparatively more different, we chose $I_L/I_P=0.3$ and the phase space is presented in Figure \ref{fig5} (c). Both chosen values of $I_L/I_P$ are highlighted in Figure \ref{fig5} (a) by the symbols with black outline.

\begin{figure}[!h]
	\centering
	\includegraphics[width=0.48\textwidth]{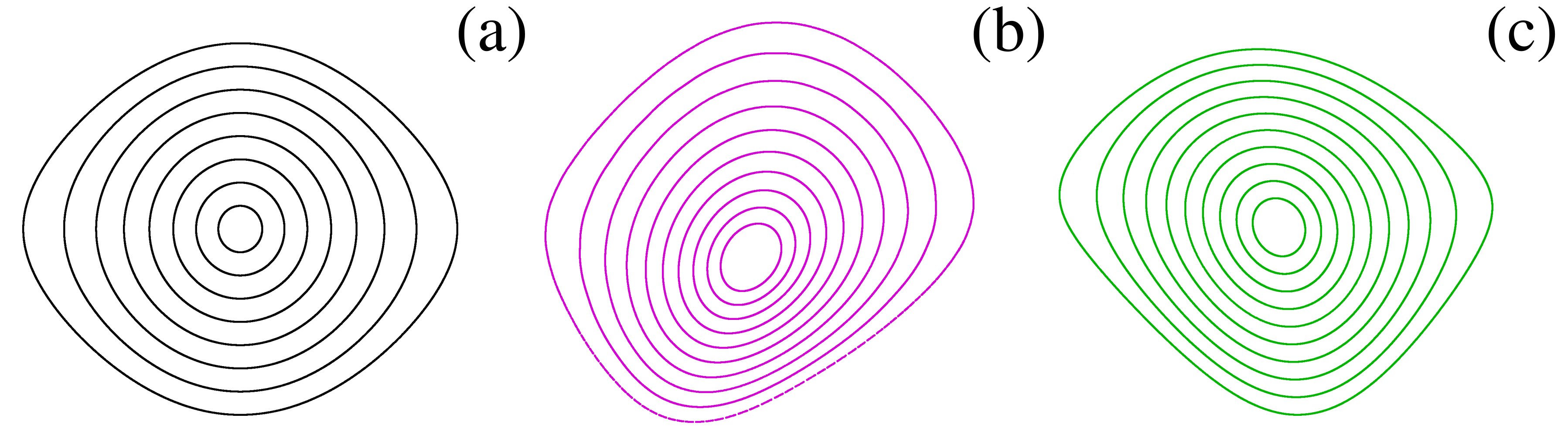}
	\caption{Comparison between the islands obtained as solution for the (a) pendulum equations and (b)-(c) for the EMT map. The island exhibited in (b) and (c) are characteristic of the lower and upper chain of islands in the extended map, respectively.}
	\label{fig6}
\end{figure}

Comparing the islands presented in Figure \ref{fig5} (b) and (c), we observe differences in size and shape. The larger size presented by the islands for $I_L/I_P=0.3$ is the result predicted by the pendulum approximation. The shape of the islands presented in Figure \ref{fig5} (b) resembles pendular islands  resulting from the pendulum equations. A different scenario is observed in Figure \ref{fig5} (c): the shape of the upper (green) islands is slightly triangular while the lower (magenta) islands have their center apparently displaced. In order to emphasize the difference, we compare directly the island obtained from the integration of the pendulum equation and the islands of the phase space in  Figure \ref{fig5} (c) (see Figure \ref{fig6}).
	
The islands observed in Figure \ref{fig6} (a) are computed by the numerical integration of pendulum equations of motion. The islands present an ellipse-shape form and are concentric around the center of the ellipse. The two chains of islands presented by the EMT map have the same period and winding number but  not the same shape. This result is expected since the map is asymmetric, what was also observed for the extended standard nontwist map \cite{mugnaine2020}. The lower islands, showed in Figure \ref{fig6} (b), are deformed ellipses: the upper arc resembles an elliptic arc whereas the lower arc is not symmetric with respect to the vertical ellipse axis, thus the islands are distorted and no longer concentric.

A similar scenario is observed for the islands of the upper chain of islands, indicated in the Figure \ref{fig6} (c). The islands are apparently concentric, but their shape is slightly triangular. Different from the lower islands discussed before, the lower ``arc" of the island is symmetric to a vertical axis. The structure of chain of lightly triangular islands is reported in Refs. \cite{corso1998,corso1997} and is related to the approximation of the two chain of islands during the reconnection process.

We conclude that the ratio of the increase in the half-width of the island does not follow the pendulum approximation since the shape of the islands becomes non-pendular as $(I_L/I_P)$ increases. Similar results were found for other values of $m$ in the same range of $I_L/I_P$, and $q_a=1.1$ for $m=1$, $q_a=2.3$ for $m=2$, and $q_a=4.7$ for $m=4$.

\section{Shearless curve}

The EMT map violates the twist condition $\partial x_{n+1}/\partial y_n \ne 0$, which results in a set of points in the phase space where the shear is null and the winding number takes on an extreme value. This set of points belongs to the shearless curve, a characteristic solution of degenerate (nontwist) systems. In this last section, we focus on the position of the shearless curve and its response to the perturbation of the ergodic limiter.

\begin{figure*}
	\centering
	\includegraphics[width=0.95\textwidth]{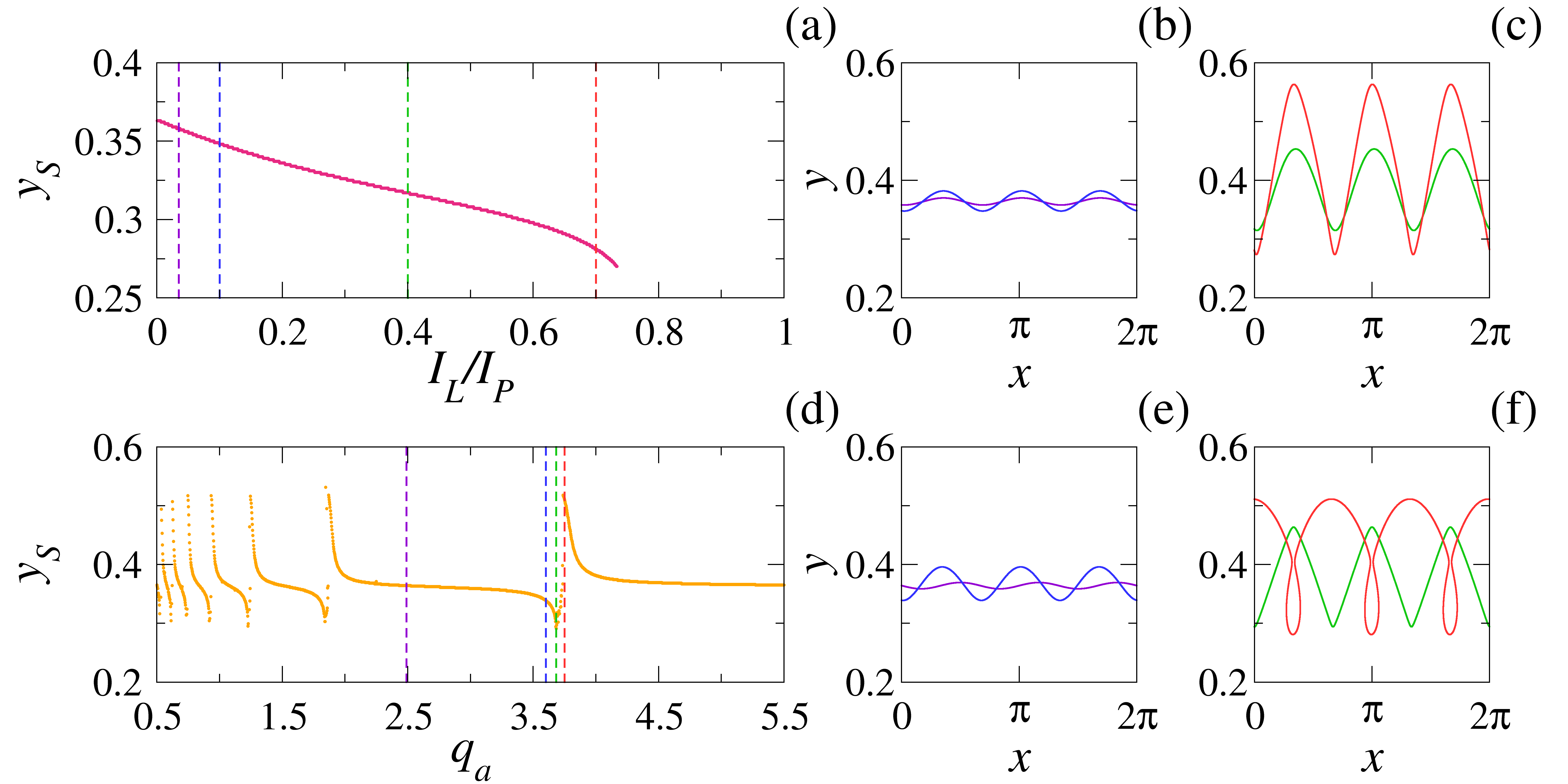}
	\caption{The effect of the control parameters on the winding number profile for the EMT map. The position $y_s$ corresponds to the point $(x_0=0,y_s)$ in the  shearless curve for different values of $I_L/I_P$ and $q_a$ for  $m=3$, (a) $q_a=3.5$ and (d) $I_L/I_P=0.1$, respectively. The respective shearless curve for the parameters indicated by the {\color{black}violet (black), blue (dark-gray), green (light-gray) and red (gray)} dashed lines in (a) and (d) are presented in (b)-(c) and (e)-(f), respectively.}
	\label{fig8}
\end{figure*}

As discussed in Section 2, the point $(x,y)=(x_0,y_S)$ in the shearless curve can be detected by the winding number profile, where $x=x_0$ is the line in which the profile is computed and $y_S$ is the coordinate where the extreme in the profile takes place. In order to analyze the effect of the perturbation of the ergodic limiter in the position of the shearless curve, we compute the winding number profile at $x_0=0$ for different values of $I_L/I_P$ and $q_a$ and identify the extreme value $y_S$. The winding number was calculated as presented in Section 2. The results related to the position of the shearless curve in $x_0=0$ are shown in Figure \ref{fig8}.

The position $y_S$ of the shearless curve, at $x=0$, for a fixed value of $q_a=3.5$ decreases monotonically with the increase of the ratio $I_L/I_P$, as observed in Figure \ref{fig8} (a). We can also observe small ``steps", indicating the position $y_S$ is fixed for a small variation in the parameter. A different scenario is observed in Figure \ref{fig8} (d), in which the position $y_S$ is calculated varying the parameter $q_a$ for a fixed $I_L/I_P=0.1$.  The values of $y_S$ have an alternating sequence of divergences, where we have a asymptotic behavior at the ``left" and ``right". This behavior repeats itself for all the analyzed range of $q_a$, where the central horizontal part increases consecutively.

In order to understand what happens with the shearless curve, we choose four points in the graphs of Figures \ref{fig8} (a) and (d) and computed the respective shearless curves. In Figure \ref{fig8} (b), we plot the shearless curve for $I_L/I_P=0.035$ and 0.1 as the violet and blue curves, respectively. In Figure \ref{fig8} (c), the respective values are $I_L/I_P=0.4$ and 0.7 for the green and red curve, respectively. We observe that increasing the ratio $I_L/I_P$ implies in the widening of ``oscillation" of each curve. With this increase, the value of $y$ at $x=0$ goes to smaller values. We also observe that the maximum and the minimums of the oscillation remain at the same points $(x,y$) for all analyzed values of $I_L/I_P$.

For Figure \ref{fig8} (d), we choose four values, namely $q_a=2.49$, 3.6, 3.6825 and 3.75, indicated by the respective violet, blue, green and red dashed lines in Figure \ref{fig8} (d) and curves in Figure \ref{fig8} (e) and (f). Comparing the curves in Figure \ref{fig8} (b) and (e), we observe a similar scenario: the shearless curves oscillate and for a larger value of the parameter, the amplitude of the oscillation is also greater. However, unlike observed for a fixed value of $q_a$ (Figure \ref{fig8} (b)), the minima and maxima of the curves in Figure \ref{fig8} (e) do not coincide. An even different scenario is observed in Figure \ref{fig8} (f). The green (red) curve corresponds to a minimum (maximum) value in the graph $y_s \times q_a$ in Figure \ref{fig8} (d). The shape of the two presented curves is different, while the green curve resembles a triangular signal, the red one appears to circulate around an island of period 3, similar to that observed in Figure \ref{fig3} (b). We can conclude that the transition between a minimum and a maximum in Figure \ref{fig8} (d) refers to the reconnection process of separatrix, reinforcing the observation that the variation of the parameter $q_a$ is responsible for the separatrices reconnection/collision process, as also observed in Figure \ref{fig3}. In Figure \ref{fig8}, we presented the results for $m=3$. As it happens for the analysis of the twin islands in Figure \ref{fig5}, the same scenario of Figure \ref{fig8} is also observed for $m=1$, 2 and 4.

\section{Conclusions}

The original Martin-Taylor map, proposed in Ref. \cite{martin1984}, can be extended as an area-preserving two dimensional map with a generic safety profile factor. {\color{black} In this paper, we considered an extended Martin-Taylor map (without the change of origin) whose phase space represents the entire radial extension of the Tokamak chamber.} We also consider a non-monotonic safety factor profile, since the map maintains its area-preserving property for any profile

Our results emphasize the relation between the parameter $m$, the number of toroidal pairs of coils segments in the limiter, and the period of the islands in the phase space. For a certain value of $m$, the map presents a $m$ number of islands, for both the monotonic and non-monotonic maps. The winding number profile exhibits a maximum point for the extended map, corresponding to the shearless curve. As expected for a map that violates the twist condition, the extended Martin-Taylor map also presents the twin island chain scenario as well as the separatrix reconnection/collision process. The latter is related to the changes on the parameter $q_a$, the value of the safety factor in the edge of the plasma.
 
{\color{black} We observe that the pendulum approximation is a good approximation for a range of the perturbation parameter, for the OMT, as long as the magnetic islands in the resonance $k=1$ are distant from the wall, $y=0$.} The chaotic sea emerges around $y=0$, therefore islands in this region experience the effect of the increase of the chaotic region for smaller values of the perturbation parameters.

For the extended map, we observe two points of resonance $y^*$, corresponding to the two islands of chains due to the non-monotonic behavior of the safety factor profile. The half-width for the islands of both chains increases monotonically with $I_L/I_P$, but they do not follow the scale rule predicted by the pendulum approximation. This results from the fact that the islands do not have a pendular form.

Since the extended map presents a non-monotonic safety factor profile, it also exhibits a shearless curve. From the position of the curve in the phase space,  we observed that, for a fixed value of $q_a$, the position of the shearless curve at $x=0$ decreases monotonically with the increase of $I_L/I_P$. This is due to the increasing in the amplitude of the oscillation presented by the curve. For a fixed value of $I_L/I_P$, the graph of the position of the shearless curve in relation to the increase of the parameter $q_a$ is similar to a ``$-\tan x$" graph. The transition between a minimum and maximum point of the graph corresponds to a reconnection process of the shearless curve in the phase space. Thus, the analysis of the position of the shearless curve for a fixed value of $I_L/I_P$ can give us an idea of where it occurs the reconnection of separatrix in the extended map.


The map presented in this paper is based on the Martin-Taylor map which has a physical interpretation in terms of toroidal plasma confinement. We showed that the extended Martin-Taylor map presents the same nontwist properties of other nontwist maps with small or no physical background. With this, the extended map is a suitable model for the study of properties and phenomenons related to degenerate systems in physical applications.

\section{Acknowledgments}
We wish to acknowledge the support of the Coordination for the Improvement of Higher Education Personnel (CAPES) under Grant No. 88887.320059/2019-00, 88881.143103/2017-01, the National Council for Scientific and Technological Development (CNPq - Grant No. 403120/2021-7, 311168/2020-5, 301019/2019-3) and Fundação de Amparo à Pesquisa do Estado de São Paulo (FAPESP) under Grant No. 2022/12736-0, 2018/03211-6. We would also like to thank the 105 Group Science (www.105groupscience.com) for fruitful discussions.

\section*{Data Availability}
The source code and data are openly available online in the Oscillations Control Group Data Repository \cite{repository}.

\section*{Appendix}
\label{append}

For the integrable scenario $p=0$, Equations (\ref{mt}) are reduced to, 
\begin{eqnarray}
    \begin{aligned}
        x_{n+1}&=x_n+\dfrac{2\pi}{q(y_{n+1})},\\
        y_{n+1}&=y_n,
    \end{aligned}
    \label{appx1}
\end{eqnarray}
with the respective equation of motion,
\begin{eqnarray}
\begin{aligned}
    \dfrac{dx}{dn}&=\dfrac{\partial H_0}{\partial y}=\dfrac{2\pi}{q(y)},\\
    \dfrac{dy}{dn}&=-\dfrac{\partial H_0}{\partial x}=0,    
    \label{appx2}
\end{aligned}
\end{eqnarray}
where $H_0$ refers to the unperturbed system. Integrating (\ref{appx2}) yields,
\begin{eqnarray}
    H_0=2\pi\int\dfrac{dy}{q(y)}.
    \label{appx3}
\end{eqnarray}
Once the perturbation is considered ($p\ne0$), the Hamiltonian equations take the form,
\begin{widetext}
\begin{eqnarray}
    \begin{aligned}
    \dfrac{dx}{dn}&=\dfrac{\partial H}{\partial y}=-\dfrac{p}{m} e^{-my}\cos(mx) \delta(n) + \dfrac{2\pi}{q(y)},\\
    \dfrac{dy}{dn}&=-\dfrac{\partial H}{\partial x}=\dfrac{1}{m} \ln \left\{ \dfrac{\cos[m x - p e^{-my}\cos (mx)]}{\cos(mx)}\right\} \delta(n).
    \label{appx4}
    \end{aligned}
\end{eqnarray}
\end{widetext}
and the respective Hamiltonian function is {\color{black}{obtained by integration of (\ref{appx4}),}}
\begin{eqnarray}
    H=H_0(y)+\dfrac{p}{m^2}\cos(mx)e^{-my}\delta(n),
    \label{appx5}
\end{eqnarray}
where $H_0$ is given by (\ref{appx3}), the logarithmic function in the last equation is expanded around $p=0$ and the terms $\mathcal{O}(p^2)$ have been dropped.


\end{document}